\documentclass[%
reprint,
superscriptaddress,
showpacs,
amsmath,amssymb,
 aps,
prb,
]{revtex4-1}

\usepackage{graphicx}
\usepackage[utf8]{inputenc}
\usepackage{amsmath,amssymb,amsfonts}
\usepackage{subfigure}
\usepackage[normalem]{ulem}
\usepackage{hyperref}
\usepackage{url}
\usepackage[normalem]{ulem}

\begin{document}
\title{Can 3D light localization be reached in ``white paint''?}
\author{Tilo Sperling}
\thanks{These authors contributed equally}
\affiliation{%
  Fachbereich Physik, Universität Konstanz, 78457 Konstanz, Germany
}%
\author{Lukas Schertel}
\thanks{These authors contributed equally}
\affiliation{%
  Fachbereich Physik, Universität Konstanz, 78457 Konstanz, Germany
}%
\affiliation{%
  Physik-Institut, Universität Zürich, Winterthurerstr 190, 8057 Zürich, Switzerland
}%
\author{Mirco Ackermann}
\thanks{These authors contributed equally}
\affiliation{%
  Physik-Institut, Universität Zürich, Winterthurerstr 190, 8057 Zürich, Switzerland
}%
\author{Geoffroy J. Aubry}
\affiliation{%
  Fachbereich Physik, Universität Konstanz, 78457 Konstanz, Germany
}%
\author{Christof M. Aegerter}
\affiliation{%
  Physik-Institut, Universität Zürich, Winterthurerstr 190, 8057 Zürich, Switzerland
}%
\author{Georg Maret}
\email{Georg.Maret@uni-konstanz.de}
\affiliation{%
  Fachbereich Physik, Universität Konstanz, 78457 Konstanz, Germany
}%

%\address{$^\dagger$ These authors contributed equally}

\begin{abstract}
When waves scatter multiple times in 3D random media, a disorder driven phase transition from diffusion to localization may occur\cite{Anderson1958,Abrahams1979}. In ``\emph{The question of classical localization: A theory of white paint?}'' P.W.Anderson suggested the possibility to observe light localization in TiO$_2$ samples\cite{Anderson1985}. We recently claimed the observation of localization effects measuring photon time of flight (ToF) distributions\cite{Stoerzer2006} and evaluating transmission profiles (TP)\cite{Sperling2013} in such TiO$_2$ samples. Here we present a careful study of the long time tail of ToF distributions and the long time behavior of the TP width for very thin samples and different turbidities that questions the localization interpretation. We further show new data that allow an alternative consistent explanation of these previous data by a fluorescence process. An adapted diffusion model including an appropriate exponential fluorescence decay accounts for the shape of the ToF distributions and the TP width. These observations question whether the strong localization regime can be reached with visible light scattering in polydisperse TiO$_2$ samples, since the disorder parameter can hardly be increased any further in such a ``white paint'' material.

\end{abstract}

\maketitle

\section{Introduction}

The prediction of a disorder induced metal-insulator phase transition made by P.W. Anderson\cite{Anderson1958} and the generalization as a wave phenomenon\cite{Anderson1972} stimulated  many theoretical and  experimental studies over more than 50 years\cite{Lagendijk2009}. Scaling theory predicts a phase transition from classical diffusion to localization to occur above two dimensions only\cite{Thouless1974,Abrahams1979}. Furthermore, advances\cite{Tiggelen2000,Skipetrov2006} in the self-consistent theory\cite{Vollhardt1980} 	 predict a position and time dependent diffusion coefficient in the localized regime. Experimental verifications of this phase transition in three dimensional highly scattering media has been a challenging task ever since. There are experimental reports with light\cite{Wiersma1997, Schuurmans1999,Stoerzer2006,Sperling2013}, ultrasound\cite{Hu2008} and ultra cold atoms\cite{Kondov2011,Jendrzejewski2012,McGehee2013,Semeghini2014}, but doubts have been raised concerning the interpretation of some of these results in terms of  localization of light waves (\hspace{1sp}\cite{Scheffold1999,Wiersma1999,Beek2012} and\cite{Scheffold2013,Maret2013}) and for ultra cold atoms\cite{Muller2014,McGehee2014}.

This controversial discussion illustrates the difficulty to carry out sufficiently complete and accurate sets of experiments, to fabricate appropriate samples and, finally, to reach a consistent interpretation of all results. In previous publications\cite{Stoerzer2006,Aegerter2006,Aegerter:07,Aegerter2007,Sperling2013,Maret2013,Sperling2014} we interpreted our measurements on strongly scattering TiO$_2$ powders as evidence for strong localization of light in three dimensions, as suggested by P.W. Anderson\cite{Anderson1985}. However, inconsistencies of recent data with the interpretation of Anderson localization led us to perform additional sensitive experimental tests. 

In this article, after presenting our experimental setups (sec.~\ref{sec:experiments}), we will show (in sec.~\ref{sec:newResults}) new multiple light scattering measurements, where deviations from the diffusion theory are observed. While these deviations were previously interpreted as Anderson localization, we now observe similar deviations in regimes of weak multiple scattering where Anderson localization should not occur.
In sec.~\ref{sec:characterization}, we characterize a weak fluorescent signal we find in all powders that were supposed to localize and show that the scaling of the deviations from diffusion with the disorder strength can be explained by a single exponential decay of this fluorescence process. We reinterpret previously published data with the help of a diffusion model including this fluorescence lifetime process (sec.~\ref{sec:fluorescence}). These experiments show that a weak fluorescent signal in some of our ``white paint'' materials was misinterpreted as a signature of strong light localization.

\section{Methods}
\label{sec:experiments}
Time delayed photons, i.e. photons that spend more time inside a multiple scattering medium than expected for classical diffusion, have previously been used to look for light localization\cite{Stoerzer2006, Sperling2013}. Our light source is a femtosecond pulsed laser system tunable between 550 and 650\,nm (further described in ref.\cite{Sperling2013}). On the detection side, we use a photo multiplier (HPM-100-40, Becker \& Hickl GmbH) for ToF measurements\cite{wolli2012,Sperling2014} and an ultra fast gateable camera system (Picostar, LaVision) for TP measurements\cite{Sperling2013,Sperling2015}.

Our samples are ``white'' powders made of TiO$_2$ nano-particles ($n_\text{anatase}\approx 2.5$ and $n_\text{rutile}\approx 2.7$\cite{TiO2index}). 
The high refractive index of the rutile phase and the low absorption in the used wavelength range make them an ideal material for scattering experiments. These samples are commercially available powders from DuPont and Sigma-Aldrich, characterized in detail in refs.\cite{StorzerDiss,FiebigDiss,wolli2012,Sperling2015}. Deviations from classical diffusion have been observed for three powders from DuPont in the rutile phase (
R700, R902 and R104)\cite{Sperling2013}. For samples with a typical filling fraction of 50\%, these white powders have an inverse turbidity $kl^*$ of 2.8, 3.4 and 3.7\cite{Sperling2015}, where $kl^*$ is defined as the product of the wave vector $k$ and the transport mean free path $l^*$. Their polydispersities range between 25-45\% with a mean diameter varying from 233 to 273\,nm.
Anatase as well as rutile powders from Sigma-Aldrich (respectively AA and AR) with  $kl^*=6.4$ and $5.2$ did not show any deviation from the diffusive behavior\cite{Berkovits1990,Aegerter2006} and are therefore used as diffusive reference samples (see fig.~\ref{fig:ToF-AA}). AA has a mean particle size of 170\,nm with 47\% polydispersity and AR has a mean particle size of 540\,nm with 37\% polydispersity\cite{Sperling2015}.

\section{Questioning the localization interpretation} 
\label{sec:newResults}

\subsection{Samples thinner than the previously inferred localization length}

\begin{figure}
\begin{center}
\subfigure[Aldrich anatase]{
            \label{fig:ToF-AA}
			\includegraphics[width=0.48\textwidth]{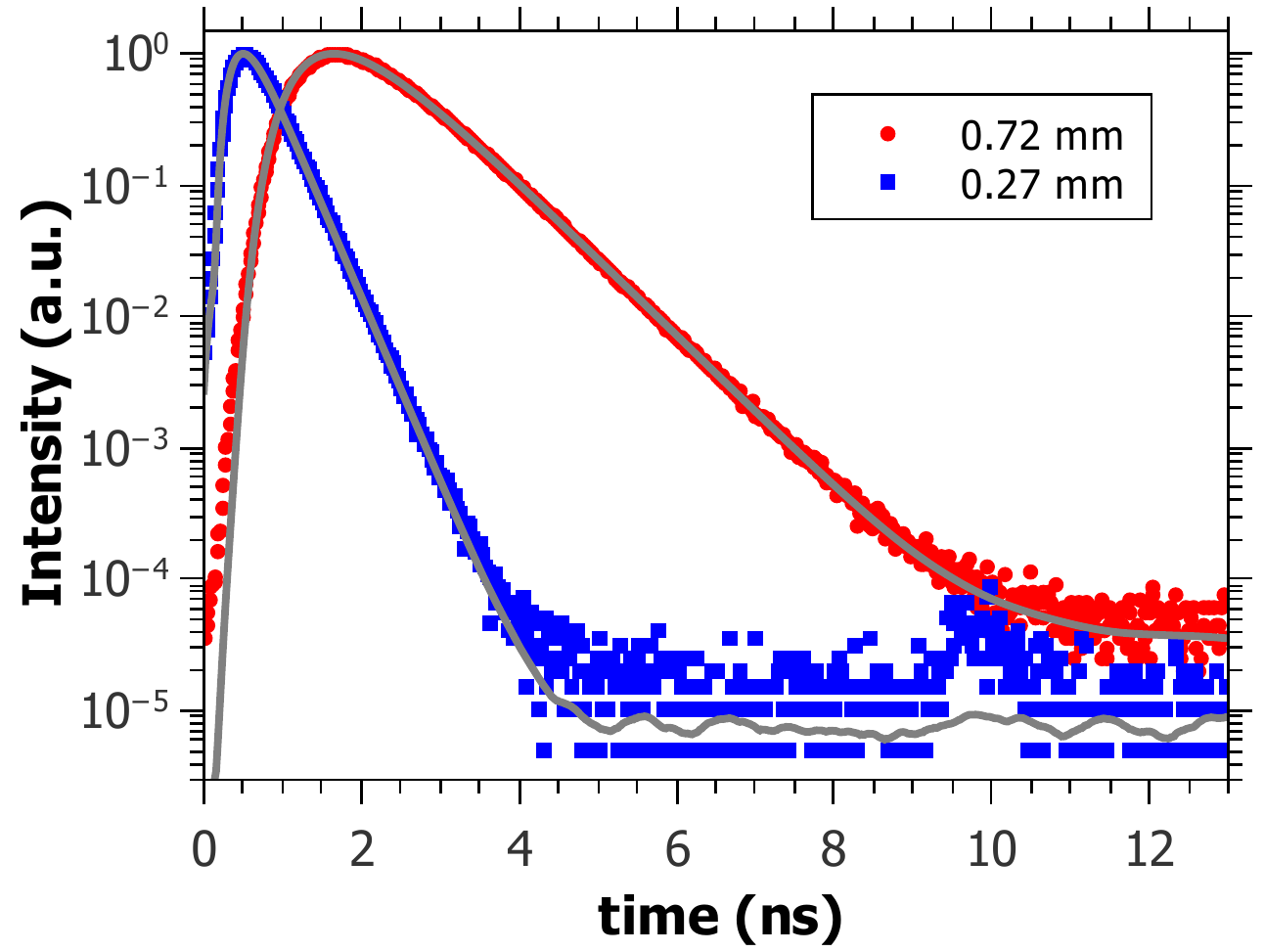}
        } 
\subfigure[R700]{
            \label{fig:ToF-R700-small}
            \includegraphics[width=0.48\textwidth]{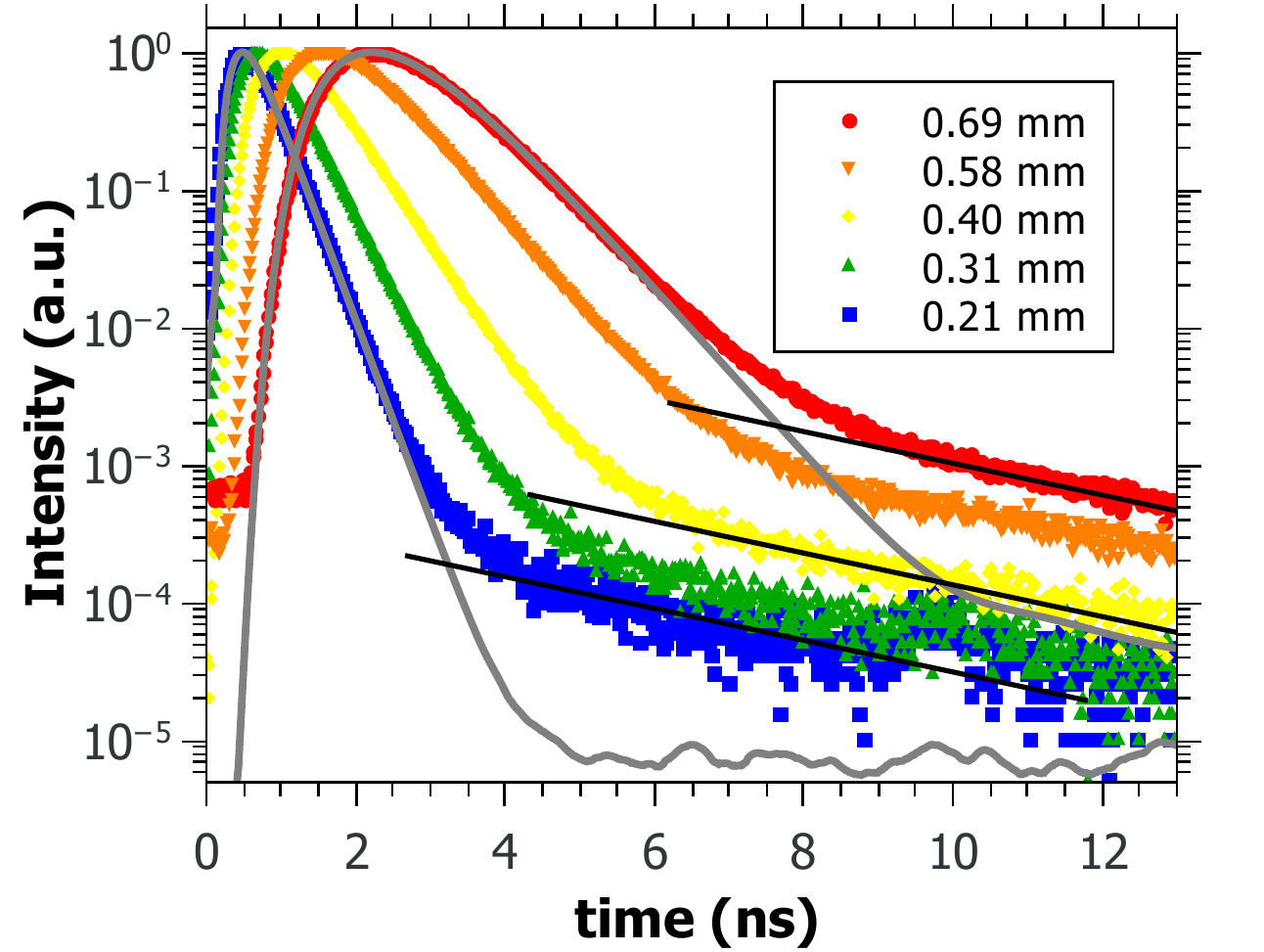}
        }
\caption{\label{fig2} (a) ToF distributions for a thin ($L=0.72$\,mm) and a very thin ($L=0.27$\,mm) AA sample are shown. The diffusion fit (solid gray lines) matches perfectly.
(b) Size dependent ToF distributions for R700 starting from a small sample size ($L=0.69$\,mm) going down to very thin sample sizes ($L=0.21$\,mm). The diffusive fit (solid gray line) does not match at long times. Black lines (guide to the eye) are shown to emphasize the exponential behavior.
The incident wavelength is 590\,nm for all measurements.}
\end{center}
\end{figure}

The theory of Anderson localization predicts the waves to be confined to a certain length scale, the localization length $\xi$. In previous experiments, this length was obtained from TP measurements, finding $\xi_{\text{R700}}=670\,\mu$m for R700\cite{Sperling2013}. It is expected that decreasing the thickness of the slab-shaped samples well below the localization length will lead to a reduction of the localization signatures since large spatial localizing modes should be significantly disturbed by boundary effects.
Figure~\ref{fig2} shows ToF distributions of AA and R700 for various thicknesses $L$.

A series of R700 samples, where the largest sample is just as thick as the earlier evaluated localization length, can be seen in figure~\ref{fig:ToF-R700-small}. For comparison, figure~\ref{fig:ToF-AA} shows two AA samples.
A diffusive fit\cite{Berkovits1990} for the thinnest and the thickest sample of each material is shown (gray lines)\footnote{The noise in the fit-curves is due to the convolution of the theory-function with a measured laser reference pulse\cite{Sperling2014}.}. For all R700 samples, even those in the $L<\xi$ regime, a clear deviation from the diffusion theory is present at long times. The photons in the long time tail seem to occur as a second exponential (black lines in fig.~\ref{fig:ToF-R700-small}) with a larger time constant. In contrast, all AA and AR samples (data not shown for AR) closely follow the diffusion theory predictions.

\subsection{Decreasing the turbidity}
\label{sec:firstSpectralMeasurements}

In 3D, Anderson localization occurs as a disorder driven phase transition.
We quantify the disorder by the turbidity $(kl^*)^{-1}$ as obtained from the width of coherent backscattering cone.
A sensitive test to check whether the long time tail originates from a second process different from localization is to strongly decrease the disorder (increase $kl^*$). In previous experiments the turbidity was varied by using different powders\cite{Stoerzer2006} and by changing the incident wavelength\cite{Sperling2013,Sperling2014}. Here, we expand the accessible range of $kl^*$ by lowering the refractive index contrast between the particles (refractive index of $n\approx 2.7$\cite{TiO2index}) and the surrounding medium, increasing thus the transport mean free path $l^*$, by replacing air ($n=1$) by agarose gel ($n\approx1.33$) as a surrounding medium.
 
\begin{figure}
\begin{center}
\subfigure[Time of flight: R700 in agarose]{
            \label{fig:ToF-R700-agarose}
            \includegraphics[width=0.48\textwidth]{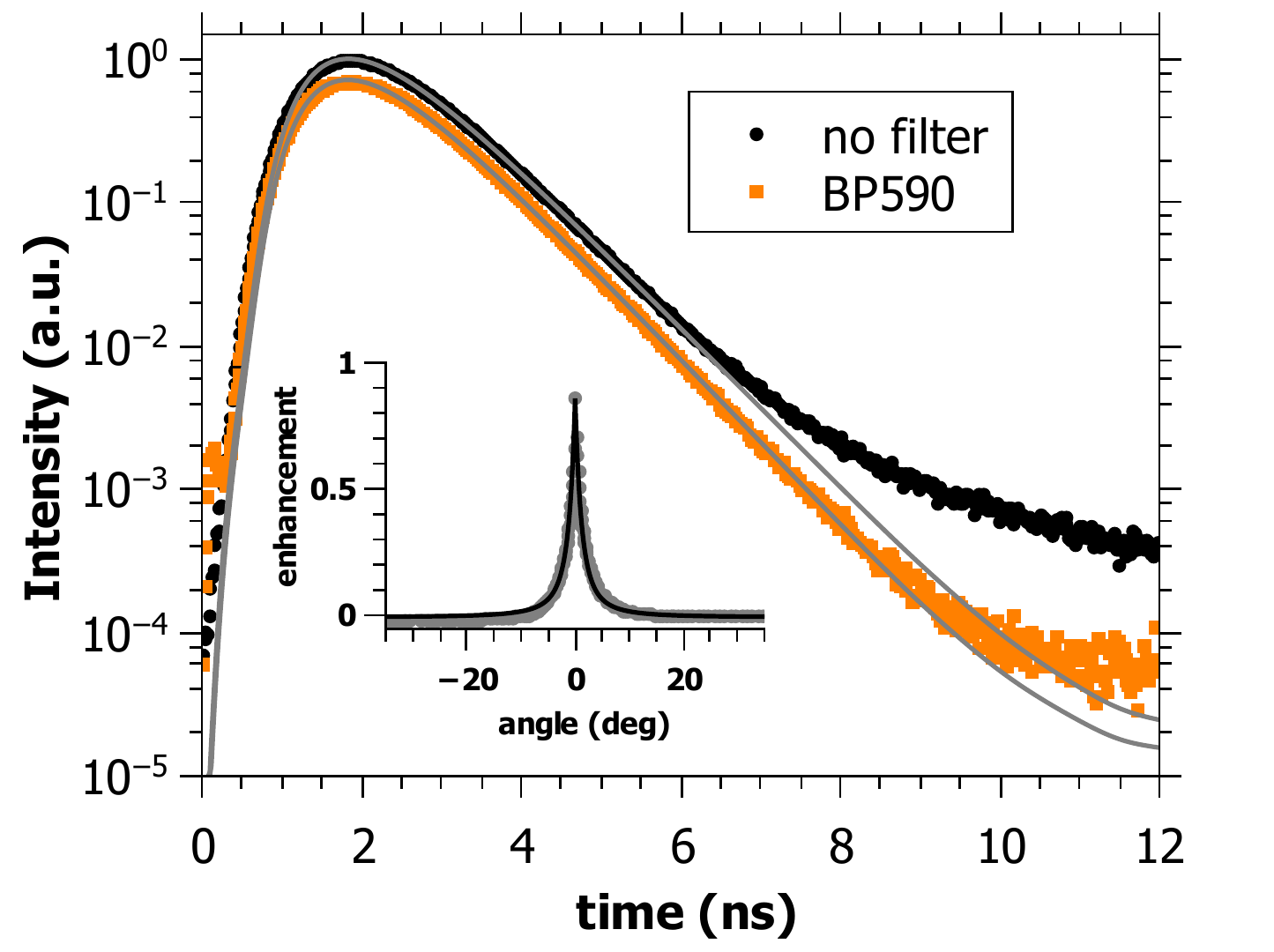}
        }
\subfigure[Transmission profile width: R700 in H$_2$O]{
            \label{fig:TP-R700-H2O}
			\includegraphics[width=0.48\textwidth]{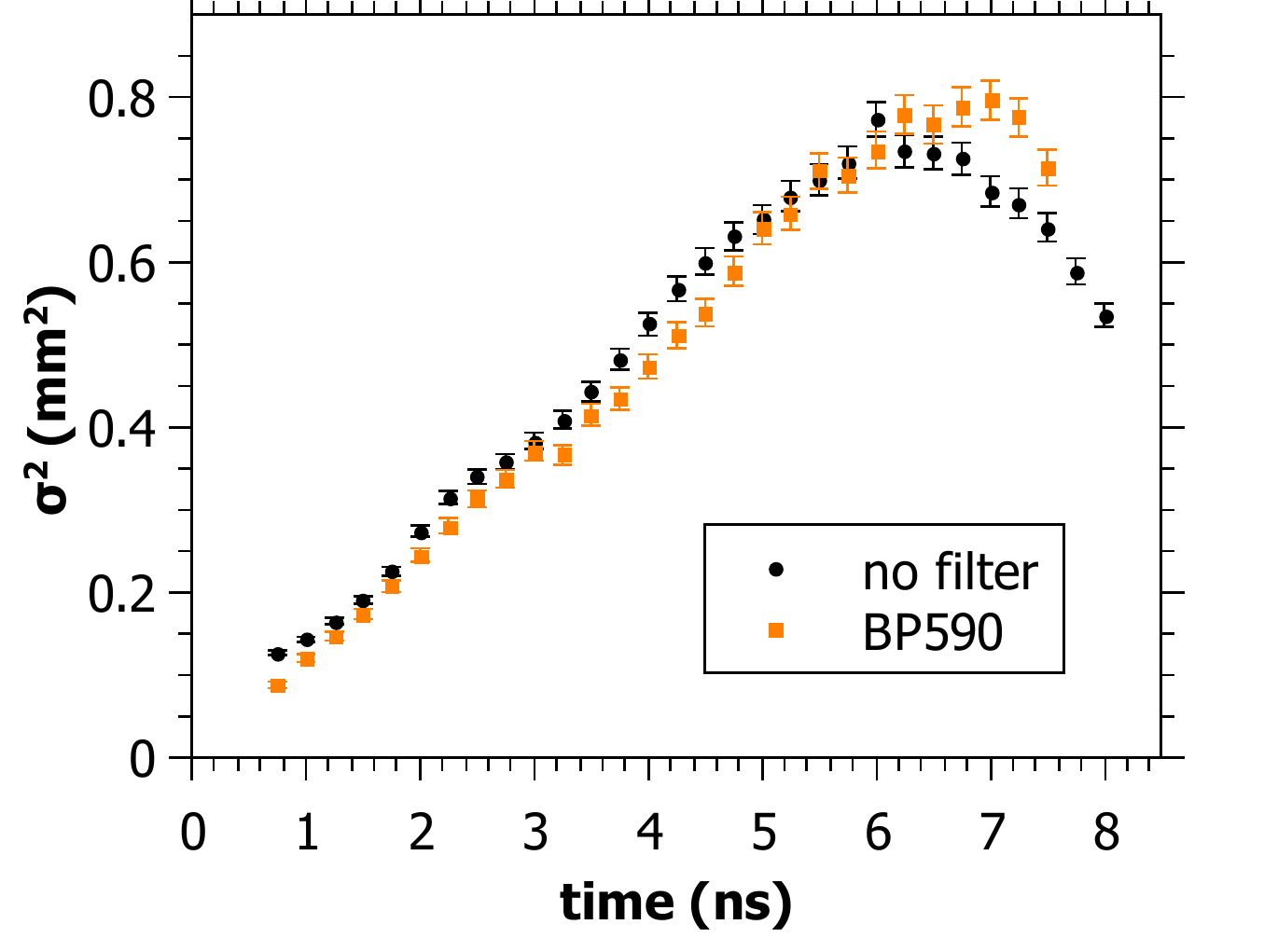}
        } 
\caption{\label{fig1} ToF distributions (a) and TP widths (b) for a fixed incident wavelength $\lambda_{\text{inc}}=590$~nm with (orange squares) and without (black dots) bandpass filter set to the incident wavelength (FWHM 10\,nm).
In (a) a sample ($L=2.3\pm 0.1$\,mm) of R700 solved in agarose gel was used. The gray lines show diffusion fits for both curves.
Inset: backscattering cone used to determine the mean free path ($l^*=1.0\pm0.1\,\mu$m)\cite{Gross2007}. In (b) R700 powder was solved in H$_2$O (ratio $1:1$). The profile width was determined following ref.\cite{Sperling2013}.}
\end{center}
\end{figure}

Figure \ref{fig:ToF-R700-agarose} shows a ToF distribution (no filter, black curve) of R700 surrounded by agarose gel.
Coherent backscattering\cite{Fiebig2008,Gross2007} was used to quantify $kl^*\approx 10$. Measuring the same sample with a 590\,nm bandpass filter  (10\,nm FWHM; BP590, orange squares in fig.~\ref{fig:ToF-R700-agarose}), as described in ref.\cite{Sperling2014}, allows us to probe the light transmitted at the incident wavelength.
Diffusive fits for both measurements are plotted in gray.
The ToF with filter follows the expected distribution for diffusive transport\cite{Berkovits1990} better than the non filtered one which shows a much more pronounced upturn for the long time tail.
Thus this long time tail must have been wavelength shifted, and is unlikely to be caused by localization since $kl^* \approx 10$ should be far in the diffusive regime. The same measurement was also performed with water or glycerol as surrounding medium, leading to the same result.

Similarly, signs of localization are tested with the TP method\cite{Sperling2013} by suspending R700 in water and evaluating the transmission profile width with and without the 590\,nm bandpass filter (see fig.~\ref{fig:TP-R700-H2O}).
The width of the transmitted profile should show a linear increase for a diffusive sample\cite{Cherroret2010}.
The data without filter (black dots) show a deviation from the linear diffusive increase of the width at long times. This deviation occurs similar to the one observed in ref.\cite{Sperling2013}, but in a higher $kl^*$ regime. The same results were obtained for the TP with glycerol as surrounding medium.
Note that in both ToF and TP with a bandpass filter, deviations from classical diffusion can be observed\cite{Sperling2014}. However, the filter has a FWHM bandwidth of 10\,nm and thus some wavelength shifted light can still pass to the detector.

In conclusion, measuring ToF's and TP's, we observed wavelength shifted photons leading to kinks at long times, which were earlier interpreted as localization signatures.
These observations are now present in a regime of low turbidity where no localization effects are expected.

\subsection{Static transmission data}

In diffusive ($kl^*\gg 1$), sufficiently thick ($L\gg l^*$) and absorbing slabs, the transmission scales with $\exp{(-L/L_\mathrm{a})}$, with $L_\mathrm{a}$ the macroscopic absorption length. In contrast, in the localization regime the total transmission is dominated by the localization length and is proportional to $\exp{(-L/\xi)}$ on top of absorption\cite{Scheffold1999}.
In early experiments, indications of localization were found in static transmission measurements performed on slabs of R700\cite{Aegerter2006,Aegerter:07}. The static transmission data could not be explained by absorption only.
Deviations were found to be in accord with the inferred localization length extracted from localization fits (\hspace{1sp}\cite{Berkovits1990}, eq.~2 in\cite{Aegerter2006}).

\begin{figure}
\begin{center}
\includegraphics[width=0.48\textwidth]{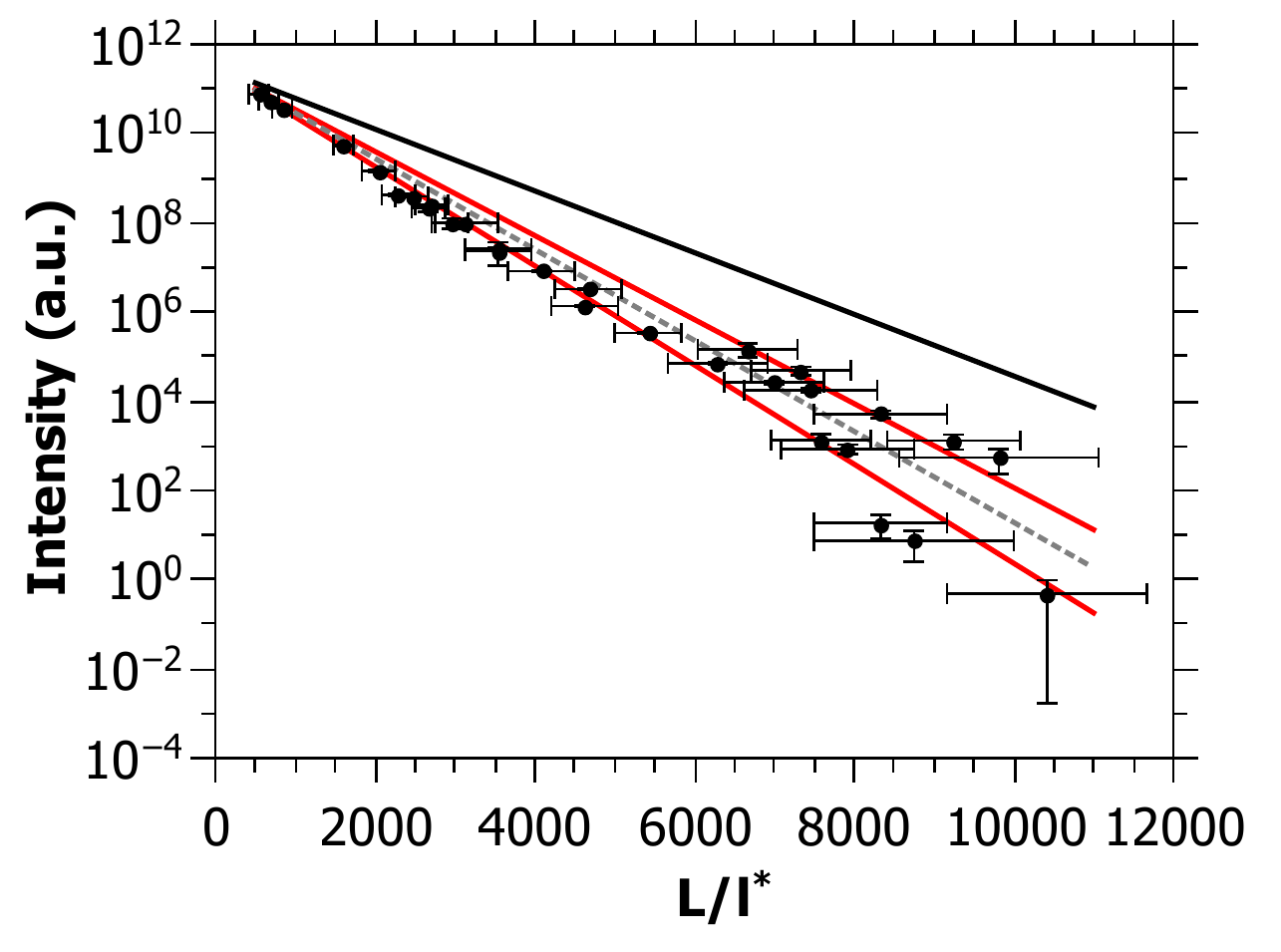}
\caption{\label{fig:transmitedI} Total transmission of a R700 sample as a function of sample thickness $L$ normalized to $l^*$. Same data as in ref.\cite{Aegerter:07}. The gray dotted line shows an exponential according to the absorption length $L_\mathrm{a}=106.5\,\mu$m with an error of $\pm 8.8\,\mu$m (red lines) obtained by diffusion fits to the ToF data of fig.~\ref{fig:ToF-R700-large}. The absorption decay can explain the data, without assuming localization effects. The black line shows an exponential according to $L_\mathrm{a}=157\,\mu$m, as obtained in ref.\cite{Aegerter2006}.}
\end{center}
\end{figure}

At this time the absorption length was obtained as a result of this localization fit on the ToF distributions, and gave a result of $L_\mathrm{a}=157\,\mu$m (\hspace{1sp}\cite{Aegerter2006}, see black line in fig.~\ref{fig:transmitedI}).
Extracting the absorption length by fitting only the diffusive part of R700 ToF distributions by diffusion theory yields a smaller average absorption length of $L_\mathrm{a}=106.5\pm8.8\,\mu$m than the one obtained by the localization fit\footnote{Despite better data evaluation, we now use a diffusion fit instead of a localization fit. It turned out that the diffusion fit results in reliable absorption times for all samples of one powder, whereas the localization fit did match the data poorly and produced scattered (and quite different) absorption times.}.
This new absorption length value allows us to re-interpret the static transmission data of ref.\cite{Aegerter2006,Aegerter:07}. The exponential decay is now explainable by absorption alone (gray dotted line in fig.~\ref{fig:transmitedI}).

Similarly, earlier claims of light localization in 3D by Wiersma et al. in the transmission coefficient\cite{Wiersma1997} could also be explained by absorption\cite{Scheffold1999, Beek2012}. The same exponential signatures of localized light and absorbed light in static transmission data make it very difficult to distinguish these effects and should be handled with care. A clear data analysis can only be guaranteed by an absorption-free time-resolved method such as the transmission profile width measurements from Sperling et al.\cite{Sperling2013}. However figure \ref{fig:TP-R700-H2O} questions the interpretation of these data.

\section{Signs of a weak fluorescent signal}
\label{sec:characterization}

\begin{figure}
\begin{center}
\includegraphics[width=0.48\textwidth]{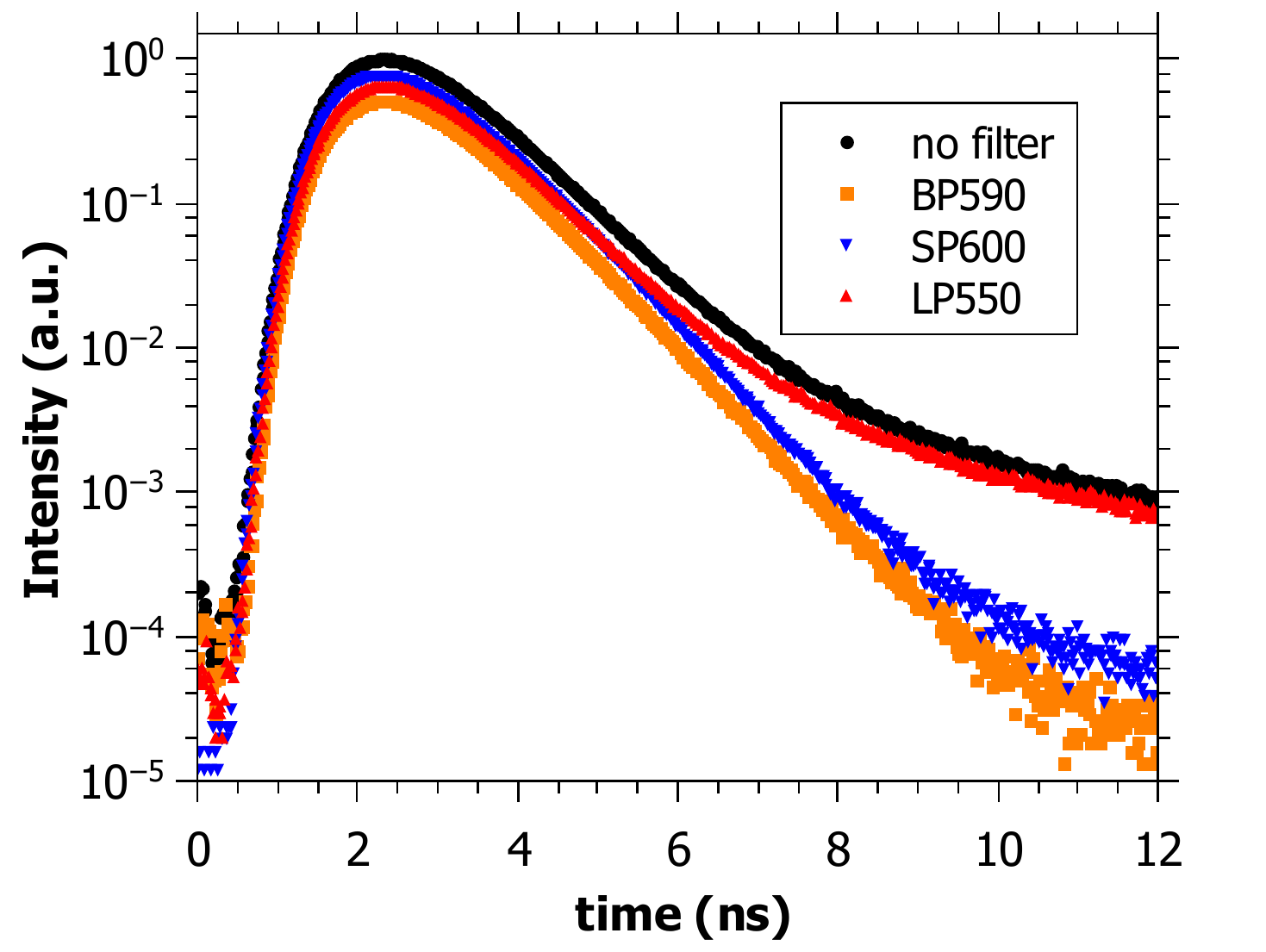}
\caption{\label{fig:ToF-R700-filters} ToF distribution of a R700 sample ($L=0.83$\,mm) measured without filter (no filter), a bandpass filter around 590\,nm (BP590), a shortpass filter 600\,nm (SP600) and a longpass filter above 550\,nm (LP550) with an incident wavelength $\lambda_{\text{inc}}=590$\,nm. The red shifted photons show a long time tail, indicating a fluorescence while the blue shifted and non shifted light behaves purely diffusively.}
\end{center}
\end{figure}

We showed in figure~\ref{fig:ToF-R700-agarose} a first crude spectral analysis of the ToF distribution for a sample consisting of R700 embedded in agarose.
This measurement indicates that the deviation in the long time tail originates from wavelength shifted photons with respect to the incident value $\lambda_{\text{inc}}=590$\,nm. In figure~\ref{fig:ToF-R700-filters} we further investigate the spectral shift of the photons in the long time tail.

ToF's of a pure R700 sample were measured using different filters, similarly to measurements performed in ref.\cite{Sperling2014}. The distribution with no filter between the sample and the detector (black dots) shows a strong upturn of the long time tail. A measurement with a bandpass filter around 590\,nm reveals that the non wavelength shifted light propagates through the sample diffusively\footnote{Up to the small effect in the late time due to the finite width of the BP590 which was already discussed in sec~\ref{sec:firstSpectralMeasurements}.} (orange squares). A measurement with a shortpass filter blocking all photons above 600\,nm (blue down triangle) highlights that the long time tail is dominated by red shifted light: the long time tail is blocked by the SP600 filter. The ToF using a longpass filter for wavelengths above 550\,nm (red up triangle) nearly matches the measurement with no filter, strengthening the observation of a signal in the red shifted region\footnote{These observations are in contrast to earlier observations published in ref.\cite{Aegerter:07}, in which a problem with the used filter can not be excluded.}. 

In the spectral study shown in fig.~\ref{fig:ToF-R700-filters}, all the photons in the long time tail occur as red shifted light. In fig. \ref{fig:ToF-R700-agarose} the long time tail occurs for wavelength shifted photons in a low scattering regime. An exponential behavior of the long time tail in ToF distributions for very thin samples is observed in fig. \ref{fig:ToF-R700-small}. All together, these observations suggest that localization claims do not hold anymore and that a lifetime process, such as fluorescence, is most likely the source of these photons.

\begin{figure}
\begin{center}
\subfigure[Emission spectra]{
            \label{fig:EmissionSpectrum}
            \includegraphics[width=0.48\textwidth]{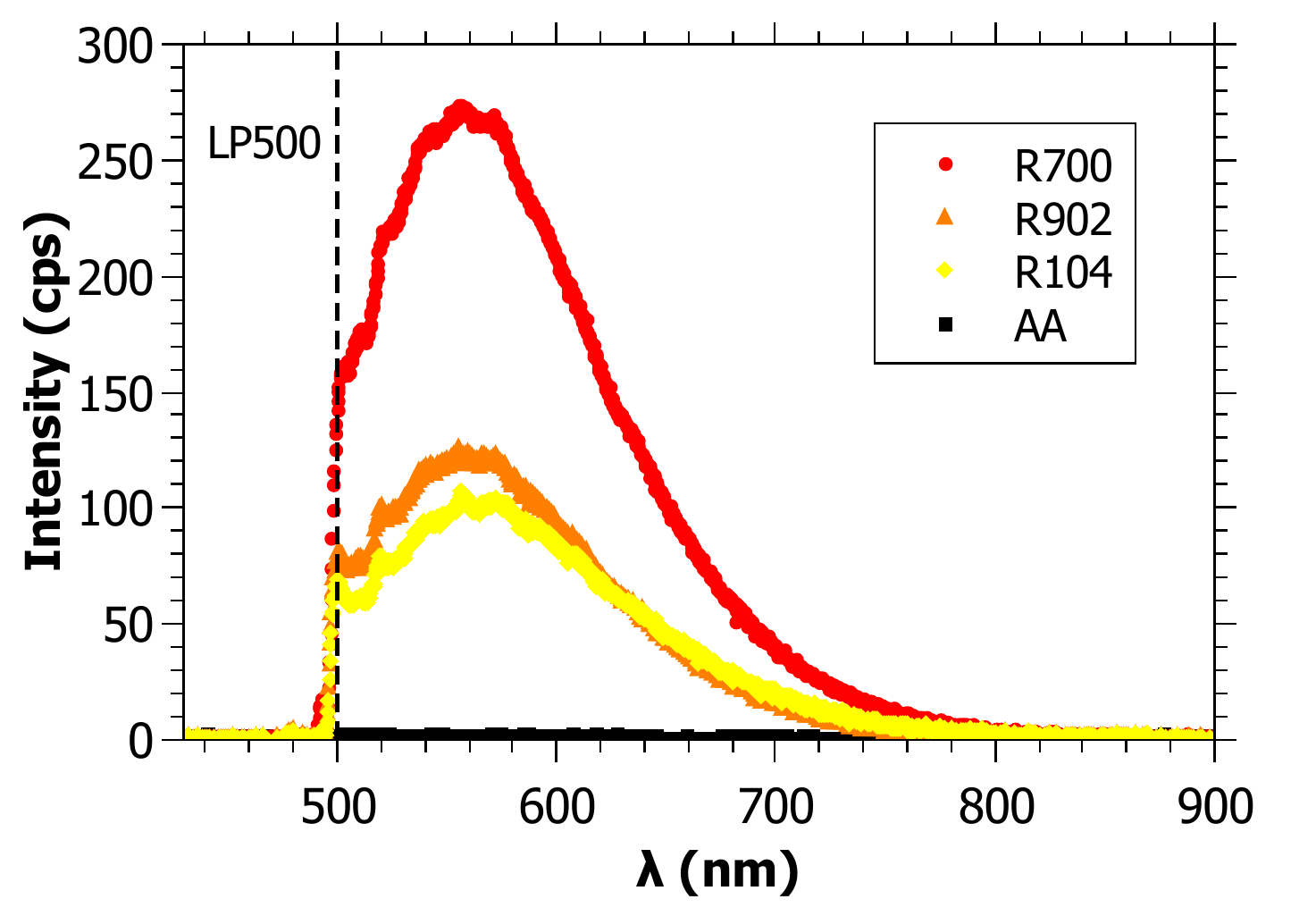}
        }
\subfigure[Lifetime measurements of R700]{
            \label{fig:lifetime}
			\includegraphics[width=0.48\textwidth]{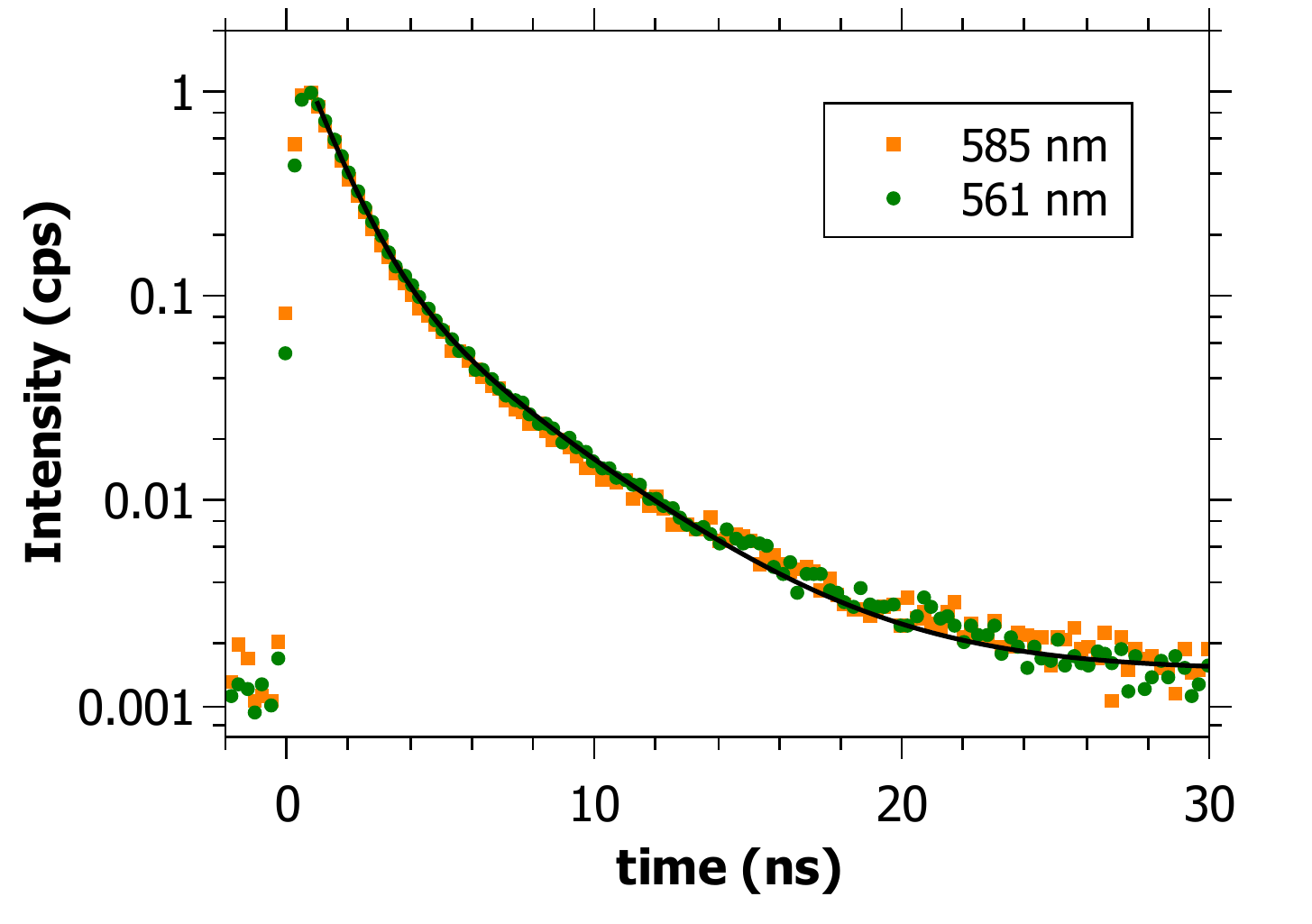}
        } 
\caption{\label{fig8} (a) Emission spectra of R700 (red dots), R902 (orange triangles), R104 (yellow diamonds) and AA (black squares) with $\lambda_{\text{inc}}=485$\,nm using a long pass 500\,nm filter and a laser power of $P=50\,\mu$W. (b) Lifetime measurements for two different incident wavelengths (585\,nm (orange squares) and 561\,nm (green dots)). The corresponding bi-exponential fit is shown as black line. The first exponent corresponds to the electronic setup response and can be ignored.}
\end{center}
\end{figure}

Thus, in order to quantify the origin of the long time tail in the ToF and the kink in the TP measurements we search for a fluorescent signal in the visible region.
The white powders are therefore spectrally analyzed in a sensitive micro-luminescence microscope setup, further described in ref.\cite{LeitenstorferSetup}.
The light source is a widely tunable pulsed ps-laser system and the detector is an EMCCD\footnote{Electron multiplying charge-coupled device} placed behind a monochromator grating. In all samples that were previously claimed to localize (R700, R902, R104) a weak fluorescent signal is observed with a broad emission in the visible range. The photo luminescence (PL) spectra of R700, R902, R104 and AA are shown in fig.~\ref{fig:EmissionSpectrum}.
All samples are excited at $\lambda_{\text{inc}}=485$\,nm with a laser power of $P=50\,\mu$W. A longpass filter 500\,nm was used to filter the scattered laser light. R700 shows the strongest signal followed by R902 and R104. For AA no fluorescent signal within the sensitivity of the setup is observed. This relative intensity dependency follows the material dependent $kl^*$-scaling of the observed deviations from diffusion in ref.\cite{ StorzerDiss,Stoerzer2006,wolli2012,Sperling2013,Sperling2015}. No fluorescent signal was detected for a rutile phase powder from Aldrich (data not shown), excluding the rutile phase to be the origin of the deviations from diffusion.

Measurements of the fluorescent lifetime of the samples were possible with a Hanbury-Brown-Twiss experiment followed by an avalanche photodiode (see ref.\cite{LeitenstorferSetup}). An average lifetime of $\tau_\text{L}=3.85\pm0.07$\,ns was extracted from exponential fits to lifetime measurements for two incident wavelengths $\lambda_{\text{inc}}=585$\,nm and $\lambda_{\text{inc}}=561$\,nm (see fig. \ref{fig:lifetime}).

\begin{figure}
\begin{center}
\subfigure[Wavelength dependency]{
            \label{fig:9a}
			\includegraphics[width=0.48\textwidth]{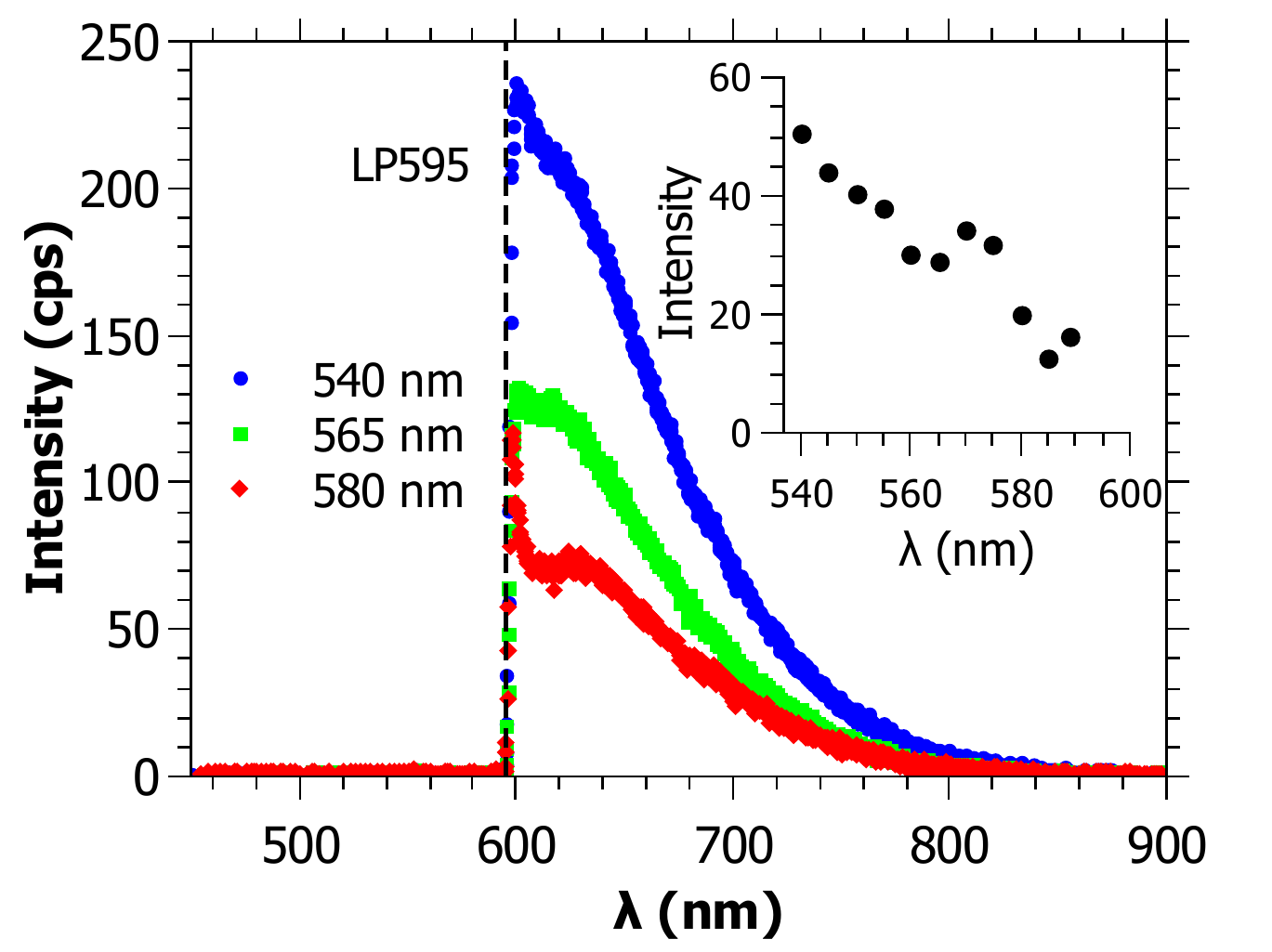}
        } 
\subfigure[Power dependency]{
            \label{fig:9b}
            \includegraphics[width=0.48\textwidth]{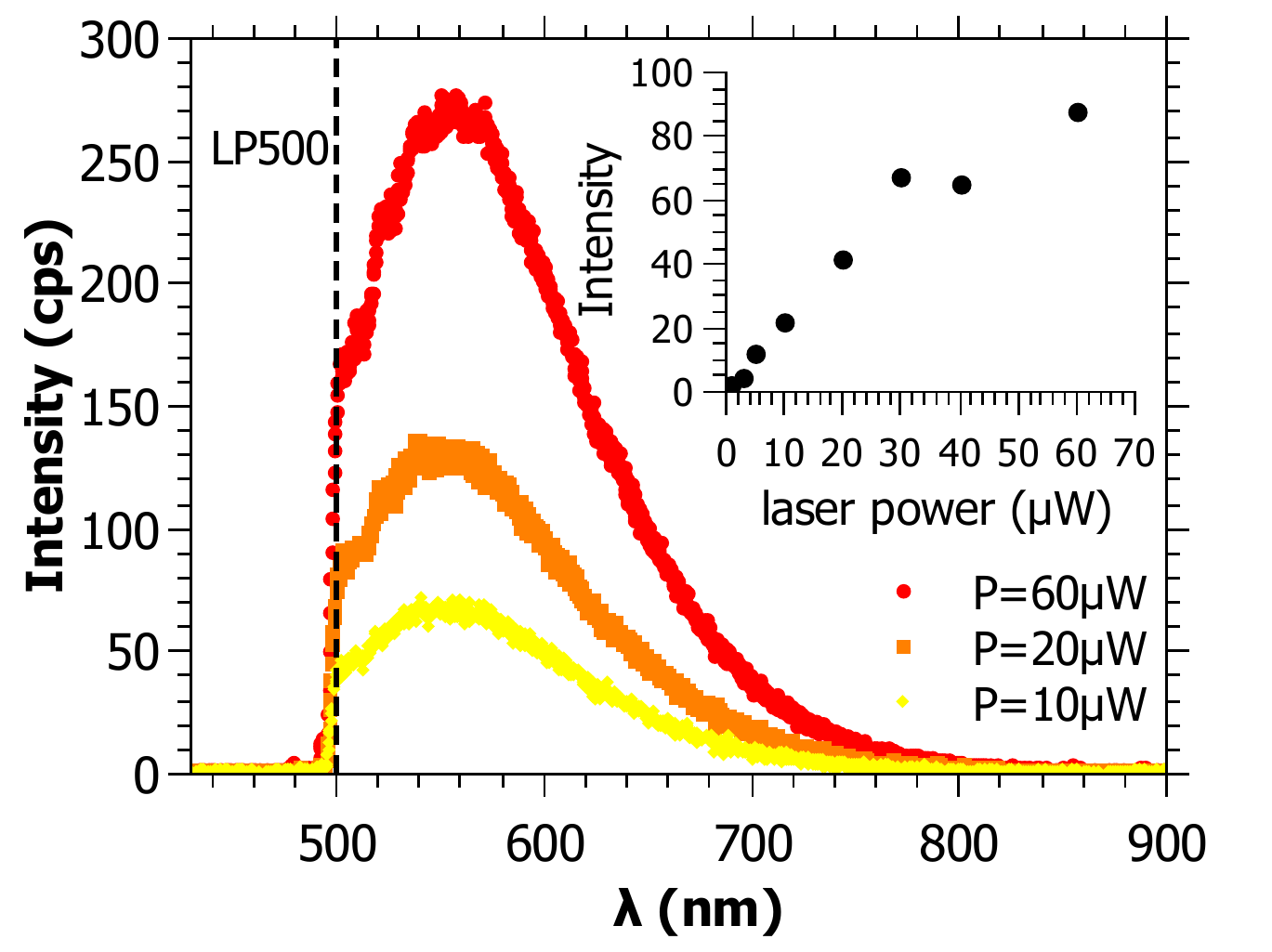}
        }

\caption{\label{fig9} (a) PL spectra for three incident wavelengths (540\,nm (blue dots), 565\,nm (green squares) and 580\,nm (red diamonds)) are shown using a longpass 595\,nm filter. Inset: integrated PLE spectra for $\lambda_\mathrm{inc}$ between 540 and 590\,nm (arbitrary unit). (b) PL spectra for three laser powers ($60\,\mu$W (red dots) $20\,\mu$W (orange squares) and $10\,\mu$W (yellow diamonds)) with $\lambda_{\text{inc}}=485$\,nm using a longpass 500\,nm filter. Inset: integrated PLE spectrum for laser power between $1\,\mu$W and $60\,\mu$W.}
\end{center}
\end{figure}

In ref.\cite{Sperling2014} an increase of the long time tail for shorter wavelength was observed, explained by the wavelength dependency of $kl^*$. Figure~\ref{fig:9a} shows PL spectra of R700 for three different incident wavelengths.
For shorter wavelength, the PL spectra increase as can be seen in the inset of fig.~\ref{fig:9a} in a range from 540\,nm to 590\,nm. This measurement explains the increase of the long time tail in ToF distributions and stronger deviations for TP's with decreasing wavelength without assuming localization effects.

The power dependent study of the fluorescence of R700 in fig. \ref{fig:9b} shows an increase with increasing incident power. The inset of fig. \ref{fig:9b} shows the power dependent integrated intensity. A slight saturation is observed as expected from fluorescence. This is in contrast to the nonlinear power dependent increase of the long time tail in ToF distributions in ref.\cite{wolli2012,Sperling2014, Sperling2015}, which have found an increase in the long time transmitted intensity with higher power. This might be due to the fact that the ToF was measured in transmission, while the fluorescence spectra were recorded in reflection. A volume of saturation for the fluorescence excitation located near the incident surface, growing with incident intensity, would explain such geometrical difference.

The occurrence of this fluorescence only in powders from DuPont led to the search of impurities in these white paint materials. An elementary analysis showed 0.2\% of carbon in R700, which could originate from organic material. In AA and AR no carbon was found, consistent with the absence of a fluorescent signal.

\section{A time delaying fluorescent diffusion model}
\label{sec:fluorescence}

To test whether the measured fluorescent signal suffices to explain our old and new data, we extended the ToF intensity distribution  $I(t)$ and the time dependent width of the TP, both known from the diffusion theory\cite{Berkovits1990,Aegerter2006,Cherroret2010}, by including a fluorescence decay.
We assume that there is an absorption rate $r_\mathrm{fl}$ by which photons traveling through the sample are absorbed. Those photons are re-emitted after a time $t_{\mathrm{d}}$ with a probability density proportional to $\exp(-t_\mathrm{d}/\tau_\mathrm{fl})$, $\tau_\mathrm{fl}$ being the fluorescence lifetime.
For a photon that, without fluorescence, takes a time $t_\mathrm{sc}$ to travel through the sample, this gives a probability density of being delayed by an additional time $t_\mathrm{d}$ due to fluorescence of
\begin{equation}
p(t_\mathrm{sc}, t_\mathrm{d}) = (1 - r_\mathrm{fl} \,t_\mathrm{sc} ) \delta(t_\mathrm{d}) + r_\mathrm{fl} \,t_\mathrm{sc}\, \frac{\exp(-t_\mathrm{d} / \tau_\mathrm{fl})}{\int_0^\infty \exp(-t' / \tau_\mathrm{fl}) \,\mathrm{d}t'}.
\end{equation}
The first term of the sum takes into account the photons that were not delayed ($t_\mathrm{d}=0$), while the second describes those which participated in a fluorescence event.
The absorption rate $r_\mathrm{fl}$ is sufficiently small that re-absorption of fluorescent photons can be neglected.

Let us recall that $I(t_\mathrm{sc})$ is the intensity of photons that arrive at time $t_\mathrm{sc}$ without fluorescence.
The intensity $I_\mathrm{fl}(t)$ for a sample showing fluorescence can now be calculated by integrating $I(t_\mathrm{sc}=t-t_\mathrm{d})$ over all delay times, weighted by the delay probability density $p(t-t_\mathrm{d}, t_\mathrm{d})$ that the photons take an additional time $t_\mathrm{d}$ due to fluorescence.
This gives a ToF distribution of
\begin{equation}
I_\mathrm{fl}(t) = \int_{0}^{t} I(t-t_\mathrm{d}) p(t-t_\mathrm{d},t_\mathrm{d}) \,\mathrm{d}t_\mathrm{d} \label{eq-tof-lum}
\end{equation}
For the TP width, $I(t)$ in eq.~\ref{eq-tof-lum} needs to be replaced by a position dependent intensity distribution $I(r, t)$ to give a 2D profile $I_\mathrm{fl}(r,t)$ from which the width can be calculated according to ref.\cite{Cherroret2010}. For fitting, the curves calculated with eq.~\ref{eq-tof-lum} were convoluted
with the time dependent detector response function for both the ToF and the TP width data.

\begin{figure}
\begin{center}
\subfigure[ToF]{
            \label{fig:ToF-R700-large}
            \includegraphics[width=0.48\textwidth]{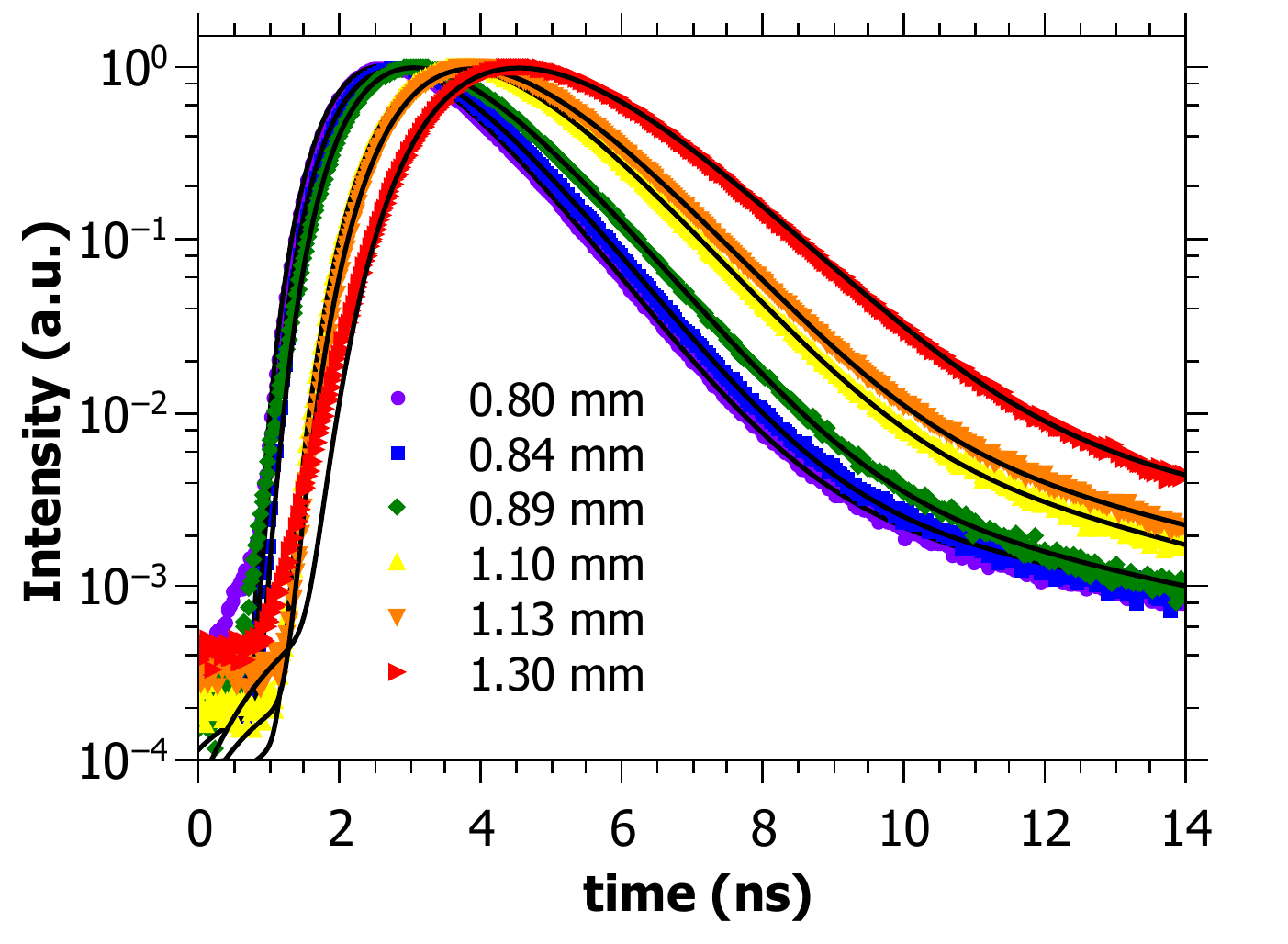}
        }
\subfigure[TP]{
            \label{fig:6b}
			\includegraphics[width=0.48\textwidth]{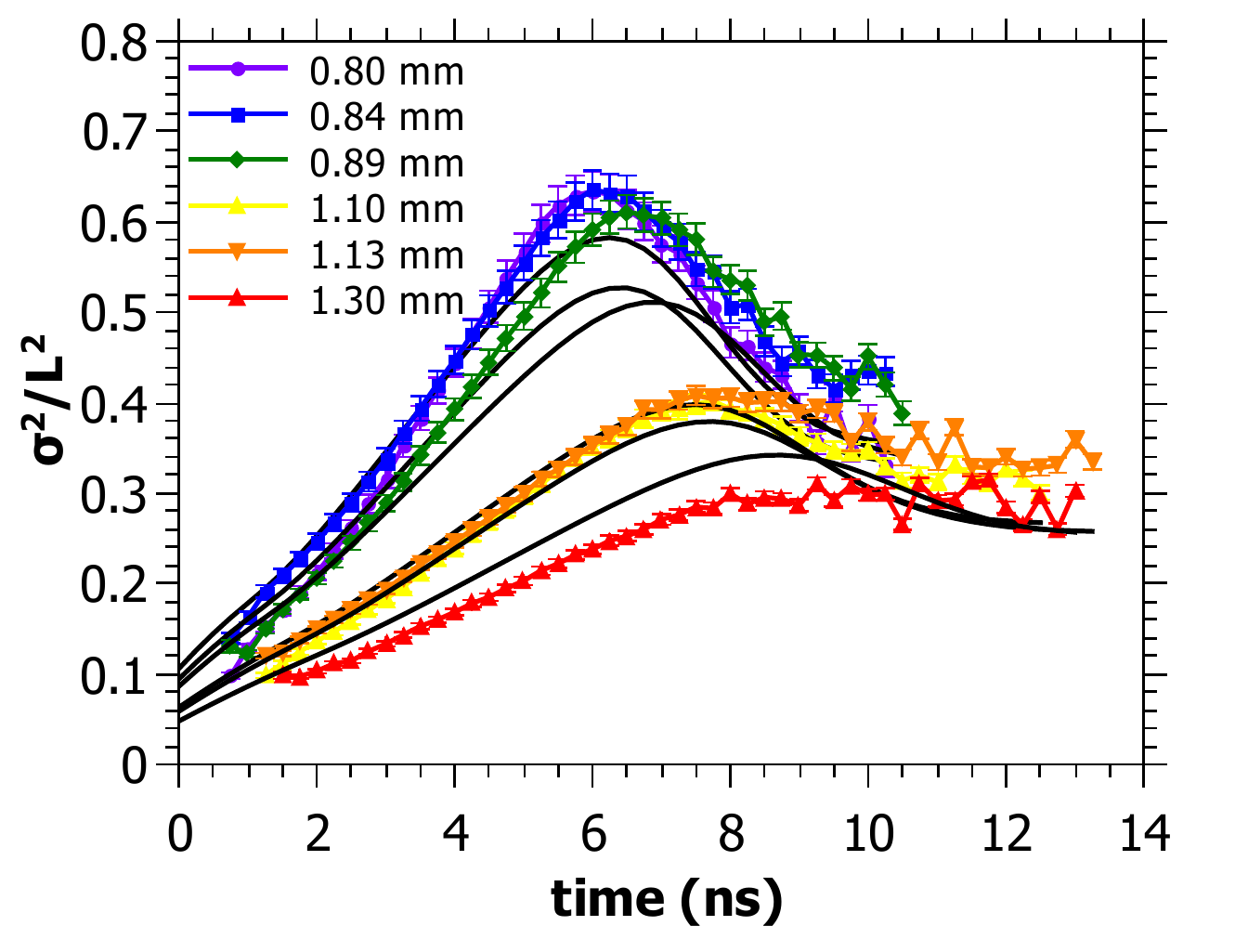}
        } 
\caption{\label{fig6} Measured ToF distributions (a) and TP widths (b) of R700 for different sample sizes $L$. ToF data taken from ref.\cite{Sperling2015}. TP data taken from ref.\cite{Sperling2013}. Diffusion fits including fluorescence from eq.~\ref{eq-tof-lum} with fixed $\tau_\mathrm{fl}=3.85$\,ns are shown as black lines.
For each sample size, the corresponding ToF and TP are fitted together, the ToF in log space and the TP in real space.
The ToF depends only on three fit parameters ($r_\mathrm{fl}$, $\tau_\mathrm{a}$ and $D$), so does the TP ($r_\mathrm{fl}$, $D$ and an offset to compensate the finite size of the illuminating beam).
}
\end{center}
\end{figure}

In figure \ref{fig6} ToF and TP measurements of R700 for different sample thickness are shown and fitted (black lines) with the extended diffusion equations that include fluorescence (eq.~\ref{eq-tof-lum}).
The corresponding ToF distribution and TP width are always fitted \emph{together} with the same set of parameters.
Each dataset was fitted with only four free parameters: the fluorescence rate $r_\mathrm{fl}$, the (usual) absorption time $\tau_\mathrm{a}$, the diffusion constant $D$ and an offset to compensate the spot size enlargement caused by the finite size of the illuminating beam in the TP measurements.
Note that the fluorescence lifetime is not fitted but set to $\tau_\mathrm{fl}=3.85$\,ns as obtained from lifetime measurements shown in fig.~\ref{fig:lifetime}. The second exponential in the ToF's is recovered very well by this decay time. In general a remarkable good agreement with the data is observed. The upturn of the long time tail in the ToF distributions can be explained by the extended theory. We are furthermore able to explain all features of the TP width measurements, in particular the thickness dependent saturation and narrowing at long times, without invoking localization effects (contrary to ref.\cite{Sperling2013}). The latter appears essentially because photons on relatively short diffusion paths (arriving at the backside of the slab at times $t< \tau_\mathrm{max}$) contribute mostly to the central part of the TP and thus, their fluorescence signal, which is delayed by the fluorescence lifetime, appears mostly in the central part of the TP. This effect gives rise to the peak in the TP width. The fits give an average fluorescence absorption rate of $r_\mathrm{fl}=0.0044\pm0.0006$\,ns$^{-1}$, an average diffusion constant of $D=11.9 \pm 0.7$\,m$^2$/s and an average absorption time of $\tau_\mathrm{a}=0.92\pm0.03$\,ns. 

\section{Conclusion}
In this article, we present new measurements that show features previously interpreted as signs of Anderson localization\cite{Stoerzer2006,Aegerter2006,Aegerter:07,Aegerter2007,Sperling2013,Maret2013,Sperling2014}, but in regimes where no localization should occur.
ToF measurements of very thin samples ($L<\xi$) still show deviations from diffusion, contrary to an expected transition to pure diffusion in the Anderson localization picture.
Furthermore, lowering the turbidity $(kl^*)^{-1}$ by changing the surrounding medium of the scattering particles does not affect the long time tail. This is also unexpected for Anderson localization since $kl^*$ is well above the expected transition value for these samples.
We were also able to show that the static transmission data of ref.\cite{Aegerter2006,Aegerter:07}, previously interpreted as a localization signature, can be actually described with absorption only, further weakening the interpretation of localization.

Besides the aforementioned observed inconsistencies, the deviations from diffusion occurred as a red shifted signal in ToF distributions. Thus we measured PL spectra for all mentioned powders in a fluorescence microscope setup, with the result that all powders earlier claimed to localize (R104, R700, R902) show a weak fluorescence signal in the visible.
Probably by chance, the samples with low $kl^*$ (reached either by using different samples or by changing the incident wavelengths) are those where the fluorescence signal is the strongest.

Finally, we performed a calculation based on diffusion theory, but including a lifetime process. This modified theory is able to fit all our data, both ToF's and TP's, with excellent agreement. A measured lifetime of the fluorescence is used in the fits as a fixed parameter and explains the second exponential decay very well.

These results strongly suggest that all deviations from pure diffusion in our ``white paint'' powders are caused by a weak fluorescence and do not originate from Anderson localization. Chemical analysis of the powders showed that they additionally contain carbon, implying the fluorescence to originate from organic impurities. However, the exact origin of the fluorescence is still unknown due to the low concentration of the impurities.

In summary, Anderson localization of light in 3D has still not been observed yet, neither in the infrared (reported in\cite{Wiersma1997}, questioned in ref.\cite{Scheffold1999} and refuted in ref.\cite{Beek2012}) nor in the visible (reported in\cite{Stoerzer2006,Sperling2013}, questioned in ref.\cite{Scheffold2013} and refuted in this article).
Although attempts have been made with higher refractive index materials (macroporous GaP --~bulk refractive index of 3.3~--~\cite{Schuurmans1999}, Ge powder --~bulk refractive index of 4~--~\cite{Rivas2001}), they all failed to reach the localization transition. Recent theoretical predictions suggest that near field effects could suppresses Anderson localization of light in 3D, but  only the case of point scatters was considered\cite{Skipetrov2014}. Also,  numerical simulations and experimental data explored these effects further, but so far in the diffusive regime only\cite{Naraghi2015}. 

Is this the end of 3D Anderson localization of light? The present reasonable answer is no: it has just not been observed yet. The quest should continue with high index ``white paint'' samples, getting rid of any fluorescent signal, and by increasing the scattering strength to currently unreached low $kl^*$ values (either in the visible or in the IR). This might be achieved by lowering the polydispersity and thus tuning the scattering to Mie-resonances in monodisperse materials.

\begin{acknowledgments}
We acknowledge support by the Deutsche Forschungsgemeinschaft (DFG), the Center for Applied Photonics (CAP), University of Konstanz and the Schweizerischer Nationalfonds (SNF). We thank the AG Leitenstorfer, and especially Denis V. Seletskiy and Florian Werschler for their most helpful cooperation regarding the PL spectra study.
We are grateful to Mengdi Chen for the elementary analysis of our powders.
We further acknowledge measurements and helpful discussions with Wolfgang Bührer.
\end{acknowledgments}

\section*{References}

\providecommand{\newblock}{}

\end{document}